\documentclass[conference,letter]{IEEEtran}
\IEEEoverridecommandlockouts

\usepackage[hidelinks]{hyperref}
\usepackage[cmex10]{amsmath}
\usepackage{amssymb,amsfonts}
\usepackage{dblfloatfix}

\usepackage[ruled,vlined]{algorithm2e}
\usepackage{graphicx}
\graphicspath{{Figures/PDF/}{Figures/PNG/}}

\usepackage{booktabs}
\usepackage{siunitx}
\usepackage[numbers,compress]{natbib}
\usepackage{texnames}
\usepackage{array}
\usepackage{bm,bbm}
\usepackage{orcidlink}

\begin{document}



\title{{A UAV-Based VNIR Hyperspectral Benchmark Dataset for Landmine and UXO Detection} \thanks{This work has been accepted and will be presented at the Indian Geoscience and Remote Sensing Symposium (InGARSS) 2025 in India and will appear in the IEEE InGARSS 2025 Proceedings.}}





\author{ 	\IEEEauthorblockN{
                Sagar Lekhak \orcidlink{0009-0009-7896-6167}$^1$,
                Emmett J. Ientilucci \orcidlink{0000-0002-3643-8245}$^1$
                Jasper Baur \orcidlink{0000-0001-7843-3939}$^2$,$^3$,
                Susmita Ghosh\orcidlink{0000-0002-1691-761X}$^4$
            } 
            \\           
	      \IEEEauthorblockA{
                \textit{$^1$Rochester Institute of Technology,}
                Chester F. Carlson Center for Imaging Science,
		      Rochester, NY 14623, USA \\ sl3088@rit.edu
                \textit{$^2$Lamont Doherty Earth Observatory,} Columbia University, New York, NY 10027, USA \\
                \textit{$^3$Demining Research Community,} New York, NY 10027, USA \\
                \textit{$^4$Jadavpur University,}
                Department of Computer Science and Engineering,
                Kolkata, West Bengal 700032, India \\
        }    }


\maketitle

\begin{abstract}


This paper introduces a novel benchmark dataset of Visible and Near-Infrared (VNIR) hyperspectral imagery acquired via an unmanned aerial vehicle (UAV) platform for landmine and unexploded ordnance (UXO) detection research. The dataset was collected over a controlled test field seeded with 143 realistic surrogate landmine and UXO targets, including surface, partially buried, and fully buried configurations. Data acquisition was performed using a Headwall Nano-Hyperspec® sensor mounted on a multi-sensor drone platform, flown at an altitude of approximately 20.6 m, capturing 270 contiguous spectral bands spanning 398–1002 nm. Radiometric calibration, orthorectification, and mosaicking were performed followed by reflectance retrieval using a two-point Empirical Line Method (ELM), with reference spectra acquired using an SVC spectroradiometer. Cross-validation against six reference objects yielded RMSE values below 1.0 and SAM values between 1° and 6° in the 400–900 nm range, demonstrating high spectral fidelity. The dataset is released alongside raw radiance cubes, GCP/AeroPoint data, and reference spectra to support reproducible research. This contribution fills a critical gap in open-access UAV-based hyperspectral data for landmine detection and offers a multi-sensor benchmark when combined with previously published drone-based electromagnetic induction (EMI) data from the same test field. Dataset Availability: \footnote{The dataset is available at: \url{https://drive.google.com/drive/folders/1h91SUWjbSjwiETcw7U9IiCBURKFGe5IJ?usp=sharing}}

\end{abstract}

\begin{IEEEkeywords}
	Hyperspectral Imaging, UAV-Based Dataset, Visible and Near-Infrared (VNIR), Landmine Detection, Humanitarian Demining, Benchmark Hyperspectral Dataset, Target Detection.
\end{IEEEkeywords}

\section{Introduction}

Hyperspectral imaging (HSI) has proven beneficial across a wide range of applications, including environmental monitoring, agriculture, mineral exploration, and military defense \cite{review_paper, review_2}. In the context of humanitarian demining, HSI offers considerable potential for detecting landmines and unexploded ordnance (UXO) by capturing the distinct spectral signatures of materials \cite{MAKKI201740}. Its capabilities for material identification and anomaly detection make it particularly effective for discriminating landmines from background clutter in complex environments \cite{landmine_detection_passive_hsi}. Airborne platforms equipped with sensors such as AVIRIS and CASI have been employed to collect VNIR–SWIR data, facilitating the identification of spectral anomalies associated with buried landmines, including disturbed soil patterns and chemical residue signatures \cite{MAKKI201740, CASI_airborne_hsi_imager_McFee1997}. Prior studies have also demonstrated the utility of HSI for simulating minefields and detecting landmines \cite{rs13050837}.

Despite these advances, the availability of publicly accessible hyperspectral datasets explicitly focused on landmine detection remains extremely limited. Due to operational security concerns and the sensitive nature of the application, most existing datasets are either unpublished or lack sufficient transparency, reproducibility, and ground-truth documentation. A notable recent effort to address the scarcity of open-access datasets is the MineInsight dataset \cite{malizia2025mineinsight}, which provides publicly available multi-sensor data for landmine research. This dataset includes visible–SWIR hyperspectral imagery, longwave infrared (LWIR) thermal data, and LiDAR measurements, acquired over a controlled test field containing 15 landmines and 20 clutter objects using an unmanned ground vehicle (UGV) platform. However, it does not include unmanned aerial vehicles (UAV) based or drone-based data to support airborne-specific use cases.

The recent proliferation of compact, low-cost, and high-resolution hyperspectral sensors has made it feasible to mount these systems on UAVs, enabling rapid and flexible surveying of large contaminated areas. A recent study \cite{data_quality_assessment} investigated the effect of UAV flight parameters on hyperspectral measurements for small landmine targets. Through spectral angle analysis, the study demonstrated that variations in drone–sensor system configurations had minimal impact on the spectral fidelity of the acquired data, highlighting the robustness and suitability of UAV-based HSI systems for demining applications.

Given the current lack of drone-based hyperspectral datasets, we present a real-world Visible and Near-Infrared (VNIR) hyperspectral dataset specifically developed for landmine detection research. The contributions of this work can be summarized as follows:

\begin{itemize}
    \item Introduction of a novel drone-based VNIR hyperspectral dataset acquired over a controlled test field containing 143 diverse realistic surrogate inert landmines and unexploded ordnance (UXO), which is radiometrically calibrated, reflectance retrieved using the Empirical Line Method (ELM), georeferenced with ground control points (GCPs) and AeroPoints, and mosaicked for comprehensive spatial coverage.
    \item Ground truth measurements (reference reflectance spectra) for all in-scene targets, and reflectance measurements of calibration panels acquired using a spectroradiometer. Accompanied by raw radiance flight-line cubes, and precise GCP and AeroPoint coordinates.
    \item Detailed documentation of the processing workflow, reflectance retrieval methods, and validation metrics to facilitate transparency, reproducibility, and ease of use by the research community.
\end{itemize}

To the best of our knowledge, there is no existing dataset that offers these features. This contribution aims to facilitate reproducible research, benchmarking, and advancement of spectral analysis and target detection algorithms for both the landmine detection community and the wider humanitarian demining research field.



\section{Test Site Description and Dataset Overview}
The dataset forms part of a larger data collection initiative led by the non-profit organization Demining Research Community, in collaboration with the Rochester Institute of Technology (RIT), aimed at advancing research in landmine detection.

\begin{figure}[t]
    \centering
    \includegraphics[width=0.85\linewidth]{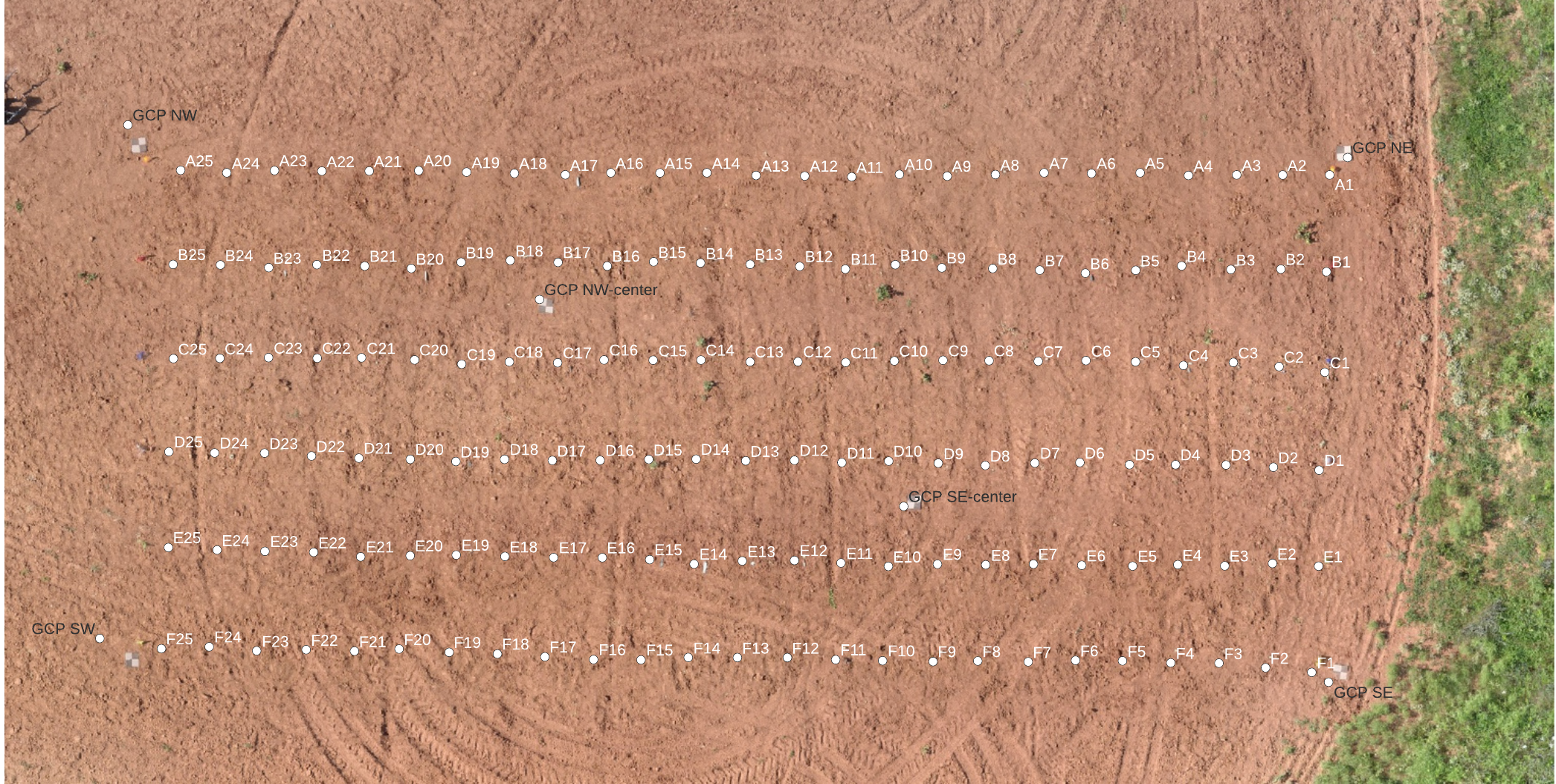}
    \caption{Pre-burial test field with target locations on bare soil in 2023. Each marker indicates a precisely geolocated target..}
    \label{fig:2023}
\end{figure}

\begin{figure}[t]
    \centering
    \includegraphics[width=0.93\linewidth]{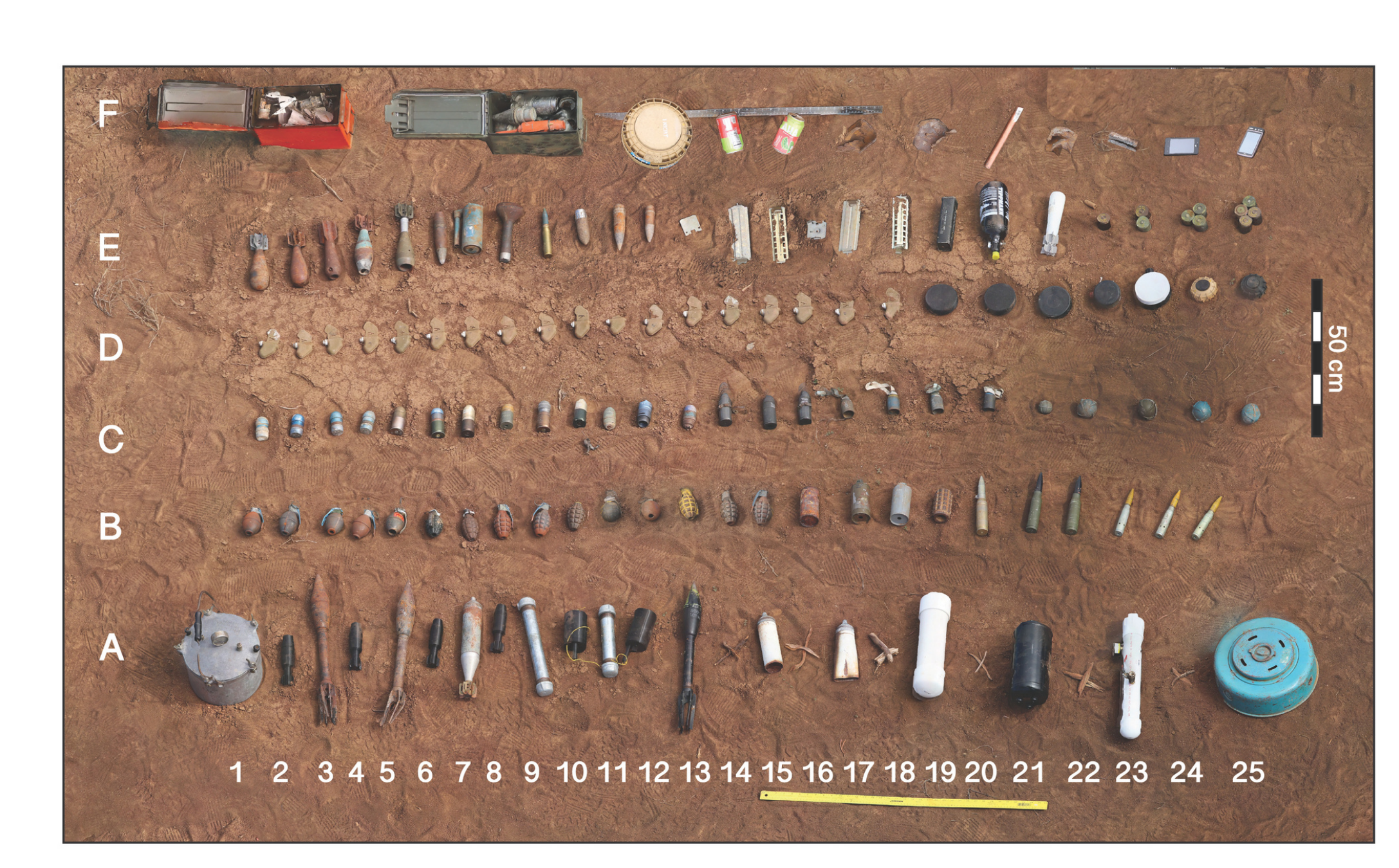}
    \caption{Condensed overview of all target types deployed in the test field pre-burial in 2023 \cite{Baur2023}.}
    \label{fig:med_res}
\end{figure}

In June 2023, the Demining Research Community partnered with the Global Consortium for Explosive Hazard Mitigation at Oklahoma State University (OSU) to seed the test site with approximately 143 diverse inert landmine objects, including inert landmines, submunitions, unexploded ordnance (UXOs), and improvised explosive devices (IEDs) \cite{Baur2023}. These objects were deployed on the surface, half-buried, and buried at varying depths and orientations to simulate realistic field conditions. A detailed description of the test site location and the types of objects deployed can be found in \cite{Baur2023,lekhak_rs}. A pre-burial image of the test site along with the precise locations of seeded objects is shown in Fig.\ref{fig:2023}. Each target’s location was geolocated at the time of manual deployment using a GPS system to ensure accurate ground truth. A condensed overview of the scene depicting all deployed mine types is provided in Fig. \ref{fig:med_res}.

Approximately one year later, in June 2024, Visible and Near-Infrared (VNIR) hyperspectral data was acquired to emulate real-world field conditions, including vegetation growth and environmental weathering. Over this period, vegetation began to obscure some surface mines, several objects were flooded by rainwater, and some buried objects reappeared on the surface, reflecting dynamic environmental effects. A time-series electromagnetic induction (EMI) dataset, collected using a drone-mounted metal detector over the same test field, was previously described in \cite{lekhak_rs}. The current work complements that effort by introducing a VNIR hyperspectral modality for optical sensing and landmine detection. Together, these datasets form a multi-sensor benchmark for landmine and UXO detection research.

The VNIR hyperspectral data was collected using the Headwall Nano-Hyperspec® sensor mounted on a multi-sensor drone (RIT’s customized Matrice 600 Pro, model MX-1) \cite{D_Kaputa}, flown at an altitude of approximately 20.62 meters. The sensor is a 640 × 1 line scanner capturing 270 contiguous spectral bands spanning the 398–1002 nm wavelength range, with an average spectral resolution of approximately 2.2 nm per band. Data acquisition was completed using 32 UAV flight lines to ensure full spatial coverage of the survey area.

\section{Data Processing}
This section describes the processing workflow from raw VNIR hyperspectral data to surface reflectance retrieval, including radiometric calibration, orthorectification, mosaicking, and empirical line correction.

The raw digital number (DN) data from each of the 32 VNIR flight lines were converted to radiance using sensor calibration coefficients provided by Headwall Photonics. Each flight line was then orthorectified using the MX-1's Applanix GPS/IMU system within Headwall’s \textit{SpectralView} software, resulting in 32 radiance cubes. These were subsequently mosaicked into a single VNIR radiance orthomosaic of the test field using \textit{ENVI Classic 5.7}.

The radiance mosaic was converted to surface reflectance using a two-point Empirical Line Method (ELM), employing the average radiance values from the light gray and black calibration panels. Prior to data acquisition, seven calibration panels with known reflectance properties were deployed near the scene. Their reflectance values were measured using a Spectra Vista Corporation (SVC) spectroradiometer and served as ground truth for ELM-based reflectance retrievals.
 
The ELM establishes a linear relationship between the measured at-sensor radiance and known surface reflectance from in-scene calibration panels. This transformation is applied independently for each spectral band to correct for atmospheric and sensor-related effects, yielding physically meaningful surface reflectance spectra. The general form of the ELM equation is:

\begin{equation}
    R = a \cdot L + b
\end{equation}

where \( R \) is the estimated surface reflectance, \( L \) is the measured at-sensor radiance, and \( a \) and \( b \) are the slope and intercept coefficients derived from the calibration panels with known reflectance. These coefficients are calculated for each spectral band using the paired radiance and reflectance values of the selected panels.

\begin{figure}[t]
    \centering
    \includegraphics[width=1\linewidth]{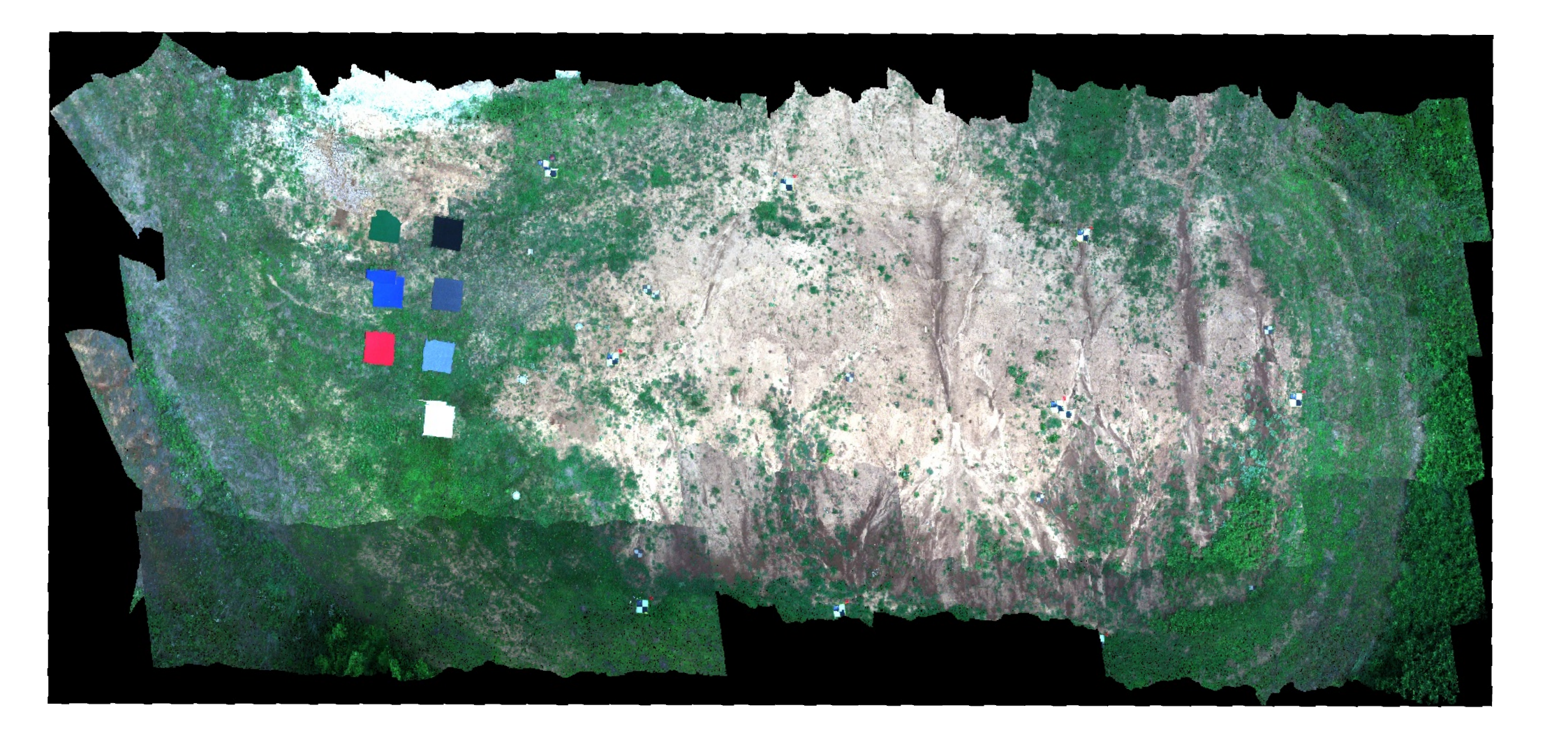}
    \caption{RGB composite of the final ELM-retrieved and georeferenced VNIR hyperspectral dataset.}
    \label{fig:elm_vnir}
\end{figure}

Afterwards, the dataset was georeferenced using the Ground Control Points (GCPs) in \textit{QGIS} software to ensure precise localization of targets within the scene. Several GCPs and AeroPoints were deployed across the test field prior to data collect to improve georeferencing accuracy. The final processed VNIR hyperspectral cube, thus obtained, has dimensions 3123×6631×272, representing spatial dimensions (lines × samples) and 272 spectral bands. An RGB composite of the final processed VNIR data is shown in Fig.~\ref{fig:elm_vnir}.

After ELM-based reflectance retrieval, the dataset was further georeferenced in \textit{QGIS} using ground control points (GCPs) to ensure precise localization of targets within the scene. Several GCPs and AeroPoints were deployed across the test field prior to data collection to enhance georeferencing accuracy. The final processed VNIR hyperspectral data cube, obtained through this workflow, has dimensions 3123 × 6631 × 272, representing the spatial layout (lines × samples) and 272 spectral bands. An RGB composite of the final reflectance data is shown in Fig.~\ref{fig:elm_vnir}.

\section{Validation Metrics: RMSE and SAM}

To evaluate the precision of the ELM applied to the VNIR hyperspectral dataset, we employed two standard quantitative metrics: Root Mean Square Error (RMSE) and Spectral Angle Mapper (SAM).

\paragraph{Root Mean Square Error (RMSE)}

RMSE quantifies the average magnitude of the error between the retrieved reflectance spectrum and the known reference spectrum across all spectral bands. It is defined as:

\begin{equation}
\text{RMSE} = \sqrt{\frac{1}{N} \sum_{i=1}^{N} \left( R_i^{\text{retrieved}} - R_i^{\text{reference}} \right)^2}
\end{equation}

\noindent where \( R_i^{\text{retrieved}} \) and \( R_i^{\text{reference}} \) are the reflectance values at the \( i^{\text{th}} \) spectral band for the retrieved and reference spectra, respectively, and \( N \) is the total number of bands. A lower RMSE indicates better radiometric accuracy of the retrieved spectrum.

\paragraph{Spectral Angle Mapper (SAM)}
SAM measures the angle between the retrieved and reference spectral vectors in an \( N \)-dimensional space, emphasizing spectral shape similarity over absolute values. It is defined as:

\begin{equation}
\text{SAM} = \cos^{-1} \left( \frac{\sum_{i=1}^{N} R_i^{\text{retrieved}} \cdot R_i^{\text{reference}}}{\left\| R^{\text{retrieved}} \right\| \cdot \left\| R^{\text{reference}} \right\|} \right)
\end{equation}

\noindent where \( \left\| R \right\| \) denotes the Euclidean norm of the spectral vector. SAM is expressed in radians or degrees, with smaller angles signifying higher similarity between the spectral shapes of the retrieved and reference spectra.

A lower RMSE value indicates better agreement, reflecting reduced radiometric errors. In contrast, SAM calculates the spectral angle between the retrieved and reference spectra in multi-dimensional spectral space, serving as an indicator of spectral shape similarity irrespective of magnitude. Smaller SAM values correspond to higher spectral fidelity, which is especially important when assessing the preservation of material-specific spectral features after retrieval. Together, RMSE and SAM provide complementary insights: RMSE captures amplitude accuracy, while SAM assesses spectral shape integrity, making them well-suited for validating the effectiveness of the ELM applied to our dataset.

\section{Validation Results}

In this section, we assess the performance of the ELM applied to the VNIR hyperspectral dataset.

To validate the accuracy of the ELM correction, six reference objects were selected: four calibration panels (Light Gray, Medium Gray, Dark Gray, and Black) and two representative in-scene targets—PFM-1 (labeled D10) and M65Al projectile (labeled A15). Two of the calibration panels (Light Gray and Black) were used for ELM, while the remaining panels (Medium Gray and Dark Gray) and the in-scene objects were reserved for validation. Ground reference measurements for these targets were acquired using an SVC spectroradiometer, which operates over a wavelength range of 338.1 nm to 2515.1 nm, producing 986 spectral bands. For comparison with the retrieved image spectra, the SVC spectra were resampled to match the image bands from 400 nm to 1000 nm. Image spectra were computed by averaging multiple pixels selected from each target to form a representative spectrum.

\begin{figure}[t]
    \centering
    \includegraphics[width=1\linewidth]{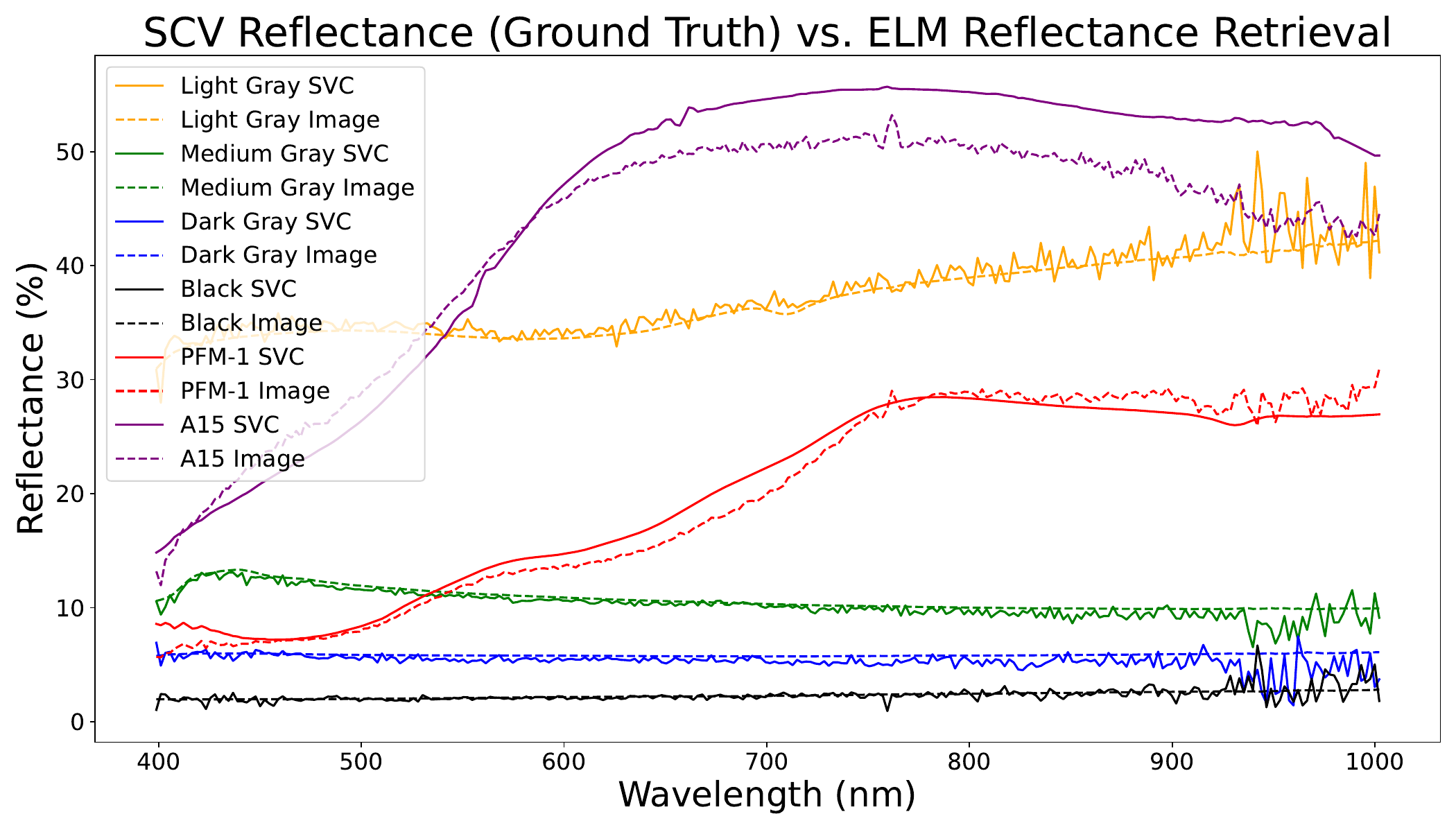}
    \caption{Comparison of reference and ELM-retrieved spectra for various in-scene materials. The label ending with ‘SVC’ refers to reference spectra measured by the SVC spectrometer, while the corresponding label with ‘Image’ indicates the ELM-retrieved spectra extracted from the hyperspectral image.}
    \label{fig:cal_panel_comparison_plot}
\end{figure}

A comparison between the SVC-measured reference spectra and the ELM-retrieved image spectra for these targets is shown in Fig.~\ref{fig:cal_panel_comparison_plot}. Visual comparison shows that the retrieved reflectance spectra closely match the reference spectra in the 400–900 nm range for the selected targets. However, beyond 900 nm, the reflectance retrievals become increasingly noisy. This is attributed to reduced sensor sensitivity, lower solar irradiance, and atmospheric water vapor absorption around the 940 nm band, which commonly affect VNIR hyperspectral sensors in this region. Additionally, the linearity assumption of the ELM becomes less reliable under low signal conditions, potentially amplifying errors beyond 900 nm.

To quantify the accuracy of the ELM, the corresponding RMSE and SAM (in degrees) values for six reference objects are summarized in Table~\ref{tab:elm_validation}. As observed, RMSE values range approximately from 0.5 to 4.5, and SAM values range from 1° to 12° when considering the full spectral range (400–1000 nm). Restricting the analysis to the 400–900 nm range improves performance, with RMSE values dropping below 1.0 and SAM values falling between 1° and 6°. Given the reflectance scale of 0–100\%, an RMSE less than 1.0 indicates very good radiometric accuracy, while a SAM of 1°–6° suggests that spectral shape is well preserved in the reflectance retrieval. The higher variation observed in the M65Al (A15) spectra can be attributed to several factors: the reference spectrometer measurements were taken on surfaces partially covered with soil and dirt; the object consists of two materials (cap and body) with potentially different spectral responses; and the exact location of spectrometer measurements on the object may have varied. In contrast, hyperspectral imagery spectra represent an average over multiple pixels, which may smooth local spectral variability.

\begin{table}[t]
   \centering
   \caption{Validation of reflectance retrievals using the ELM. }
   \label{tab:elm_validation}
   \small 
   \setlength{\tabcolsep}{3.5pt} 
   \begin{tabular}{@{}l l S[table-format=1.4] S[table-format=1.4] S[table-format=2.4] S[table-format=2.4]@{}}
       \toprule
       \textbf{Object} & \textbf{Role} & \shortstack{\textbf{RMSE}\\ \textbf{400–1000} \\ \textbf{nm}} & \shortstack{\textbf{RMSE}\\ \textbf{400–900} \\ \textbf{nm}} & \shortstack{\textbf{SAM(°)}\\ \textbf{400–1000} \\ \textbf{nm}} & \shortstack{\textbf{SAM(°)}\\ \textbf{400–900} \\ \textbf{nm}} \\
       \midrule
       Light Gray        & ELM & 1.44 & 0.83 & 1.93  & 1.06 \\
       Black             & ELM & 0.52 & 0.24 & 12.49 & 6.20 \\
       Medium Gray       & Val & 0.68 & 0.41 & 3.06  & 1.40 \\
       Dark Gray         & Val & 0.91 & 0.49 & 7.96  & 3.40 \\
       PFM-1 (D10)       & Val & 1.35 & 0.97 & 3.64  & 3.43 \\
       M65Al (A15)       & Val & 4.38 & 0.90 & 3.59  & 3.08 \\
       \bottomrule
   \end{tabular}
\end{table}

\section{Discussion}

This work represents an important step toward addressing the current lack of benchmark datasets for drone-based hyperspectral imaging (HSI) in the context of landmine detection. By releasing this dataset alongside detailed documentation of the UAV-based HSI acquisition process, reflectance retrieval methodology, and ground-truth reference measurements, we aim to empower the research community to develop and evaluate more reliable and robust spectral analysis and target detection algorithms. This resource is intended not only to support reproducible research but also to increase confidence among researchers, practitioners, and technology vendors in the practical viability of HSI for humanitarian demining.

Moreover, the availability of complementary drone-based electromagnetic induction (EMI) data from the same test field, as reported in \cite{lekhak_rs}, opens promising avenues for multi-sensor fusion research. The combination of HSI and EMI modalities, both spatially co-registered and ground-truth validated, presents a rare opportunity to develop and rigorously test advanced fusion frameworks, especially for high-risk scenarios where detection reliability is critical and false negatives are unacceptable. With the inclusion of detailed metadata and ground-truth measurements, this dataset offers a solid foundation for exploring machine learning and deep learning approaches in spectral target detection, classification, and fusion across sensing modalities. We hope this contribution fosters further innovation in spectral-based landmine detection and encourages the development of safer, more efficient demining technologies.

While the final reflectance mosaic achieves accurate geolocation for primary targets, minor distortions and radiometric inconsistencies remain in some non-critical regions due to GPS drift, drone motion, and variable illumination across flight lines. Although some AeroPoints or calibration panels appear distorted or duplicated, all surface-deployed targets remain clearly visible and correctly positioned. These limitations highlight practical challenges in UAV-based hyperspectral mapping under real-world conditions. Georeferencing individual flight lines using the provided GCPs and AeroPoint coordinates prior to mosaicking could potentially yield even finer orthomosaic alignment. By providing access to the raw flight-line data and ground control points, we enable researchers to further refine georeferencing and correction workflows as needed. Future efforts could explore advanced radiometric normalization techniques to enhance consistency, particularly in low-brightness or variably illuminated areas.

\small
\bibliographystyle{IEEEtranN}
\bibliography{references}

\end{document}